# Development of NPL Rb fountain frequency standard

*Yuri B. Ovchinnikov*

National Physical Laboratory
Teddington, Middlesex, TW11 0LW, UK
E-mail: yuri.ovchinnikov@npl.co.uk

*Abstract* — **The first characterization of the distributed cavity phase frequency shift of the NPL Rb fountain frequency standard is reported. It is shown how the symmetric part of the distributed cavity phase shift can be used to set up the vertical angle of the fountain. The estimated uncertainty of the distributed cavity phase frequency shift of $1.25 \times 10^{-16}$ reduces the total type B uncertainty of the Rb fountain down to $2.4 \times 10^{-16}$.**

## I. INTRODUCTION

The NPL Rb fountain frequency standard [1-3] is designed to have both high frequency stability and accuracy. The main advantage of Rb atoms is their low collision cross section at the low temperatures used in atomic fountains. It has been proved both theoretically [4] and experimentally [5,6] that the collision cross section of Rb atoms is about 30 times smaller than Cs atoms. This makes it possible to reduce the systematic frequency shift related to collisions between the Rb atoms. It also allows us to operate the Rb fountain with a larger number of atoms, which reduces the quantum projection noise of the standard.

One of the distinctive features of the NPL Rb fountain is the use of a Ramsey microwave cavity, with a loaded quality factor 28500, which is very close to the quality factor 32100 of an ideal copper cavity of the same size. This is intended to suppress the corresponding distributed cavity phase (DCP) frequency shifts [7-11].

In this paper we provide the results of the first characterization of the DCP frequency shift of the NPL Rb fountain frequency standard.

## II. DISTRIBUTED CAVITY PHASE SHIFT

### A. General

The DCP frequency shift [7-11] arises from the presence of phase gradients in the microwave field inside the Ramsey cavity. The atom trajectory, which is not exactly vertical, passes the Ramsey cavity in different places. The resulting frequency shift is proportional to the phase difference of the interrogation field, experienced by the atoms on their way up and down.

In the theoretical work [7] it was proposed that the transverse phase distribution of the microwave field inside a TE011 cavity could be described as a Fourier series of $\cos(m\phi)$, where $m = 0,1,2\ldots$ and $\phi$ is the azimuthal angle. The $m = 0$ term corresponds to the azimuthally symmetric phase distribution, the $m = 1$ term is related to the linear transverse gradient of the phase, and the $m = 2$ term corresponds to the quadrupolar phase variation. The higher terms are claimed to be much smaller and were neglected in the existing characterizations [9-11] of the DCP frequency shift and its power dependencies.

### B. Setting the fountain vertically

Setting the fountain vertically makes it possible to eliminate the asymmetric component of the DCP frequency shift, which provides the largest uncertainty of the DCP frequency shift in the most accurate fountains [11, 12].

The simplest way to set up the fountain vertically is to maximize the number of atoms returned to the detection region of the fountain after their launching. Possible directions of propagation of atoms in the fountain are restricted by several apertures, the smallest of which is formed by the cut-off tubes of the Ramsey cavity with inner diameter 1.6 cm. The lower cut-off tube limits the transverse velocity of the atoms reaching the detection region to a maximum value of 1 cm/s, which corresponds to an initial full divergence of the atomic cloud ~ 4.6 mrad and an average transverse kinetic energy of 0.5 μK. The atoms are initially launched vertically along the axis of the fountain with vertical velocity of 4.32 m/s. Taking into account such a large initial divergence of the atomic cloud, the precise alignment of its vertical angle is a challenging problem.

Fig. 1 and Fig. 2 show the amplitude of the detection signal at normal operation of the fountain as a function of its angle. The figures correspond to the tilting of the fountain perpendicular to or along the feeds of the Ramsey cavity (and direction of propagation of the detection laser beams).

The relative angles presented in the graphs are measured at the test platform, which is rigidly attached to the bottom of the fountain. These measurements give us the vertical angles of the fountain of $\alpha_\| = 7.7$ mrad and $\alpha_\perp = 1.8$ mrad.

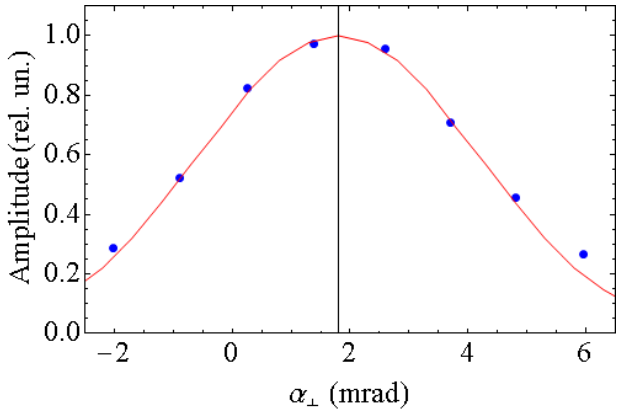

Figure 1. Amplitude of the detection signal as a function of the angle of the fountain, changed perpendicularly to the feeds of the Ramsey cavity. The fit line is the calculated dependence of the number of atoms reaching the detection region in the presence of the cut-off tubes of the Ramsey cavity for an atomic cloud with a temperature of 0.5 μK.

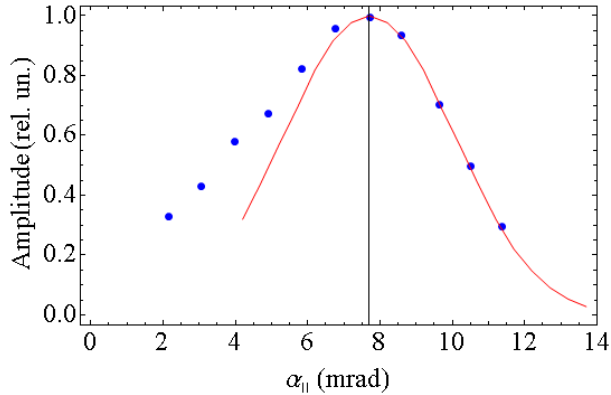

Figure 2. Amplitude of the detection signal as a function of the angle of the fountain, changed along the feeds of the Ramsey cavity. The fit line is the same as in the previous figure. The asymmetry of the experimental profile explained by the light pressure acceleration of atoms along the detection laser beam, which is directed along the feeds.

A more accurate measurement of the verticality of the fountain can be performed by measurement of the asymmetric component of the DCP frequency shift of the fountain. Here, by asymmetric DCP shift we mean the part of the shift that is not symmetric with respect to the symmetry axis of the cylindrical cavity. The main component of this shift is the m=1 term [7, 8]. In the work [9] the $\alpha_\|$ of the Cs fountain was set to zero with an uncertainty of 0.1 mrad. To achieve this, the Ramsey cavity of the tilted fountain was fed alternately from one side or the other, and the differential m=1 DCP frequency shift was measured as a function of the fountain's angle at several different amplitudes of the microwave field inside the cavity.

In our case, to set up the vertical direction of the fountain in the parallel plane, we measured the angular dependence of the asymmetric DCP frequency shift for two amplitudes of the microwave field inside the cavity, which were set near its $3\pi/2$ and $5\pi/2$ Ramsey resonances, while the cavity was excited from one side only. These microwave amplitudes were chosen because the symmetric component of the DCP frequency shift at these powers is essentially suppressed. It is important also that the cavity pulling frequency shift at these elevated powers is about 10 times smaller than at the optimal amplitude, which corresponds to the $\pi/2$ Ramsey resonance. Note, that due to the small collision cross section of the Rb atoms, its dependence on the microwave power can be completely neglected. The corresponding dependencies are presented in Fig. 3. From the independent measurements of the symmetric DCP frequency shifts (Fig. 7) and their angular dependencies in the perpendicular plane, it is concluded that their possible input to the measured relative frequency DCP shift of the fountain at these amplitudes of the field is below $10^{-15}$.

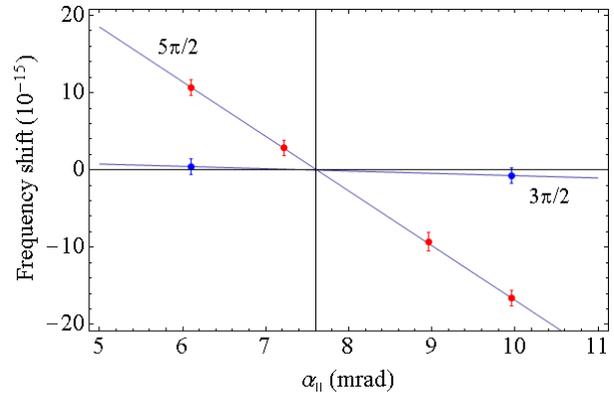

Figure 3. Angular dependence of the asymmetric DCP frequency shift for amplitudes of the microwave field inside the Ramsey cavity at $3\pi/2$ and $5\pi/2$ Ramsey resonances.

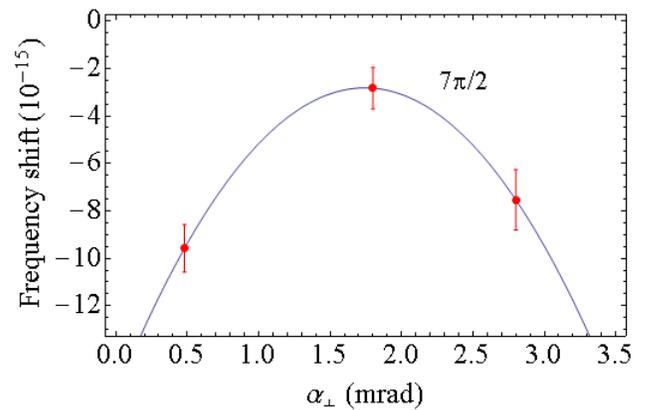

Figure 4. Angular dependence of the symmetric DCP frequency shift for the $7\pi/2$ Ramsey resonance of the microwave field.

The determination of the vertical direction of the fountain in the plane perpendicular to the feeds of the cavity cannot be done by measuring the angular dependence of the asymmetric DCP frequency shift, because in this direction it is either absent or very small. On the other hand, the symmetric DCP shift can be used to solve this problem.

We have found that the angular dependence of the symmetric DCP shift is essentially enhanced at the $7\pi/2$ and $9\pi/2$ amplitudes of the microwave field inside the Ramsey cavity. The parabolic fit of the angular dependence of the transverse symmetric DCP frequency shift on Fig. 4 gives the vertical angle of the fountain $\alpha_\perp = 1.74 \pm 0.2$ mrad.

Summarizing the results, both methods of determination of the vertical direction of the fountain give about the same resulting angles of $\alpha_\| = 7.6 \pm 0.2$ mrad and $\alpha_\perp = 1.74 \pm 0.2$ mrad.

### C. Measurement of the asymmetric DCP-shift uncertainty

The measurement of the uncertainty related to the asymmetric DCP frequency shift was done after carefully balancing the powers of the two feeds to the Ramsey cavity at the amplitude of the microwave field corresponding to a $\pi/2$ Ramsey resonance. The corresponding data on the angular dependence of this shift in the two orthogonal directions are presented in Fig. 5 and Fig. 6.

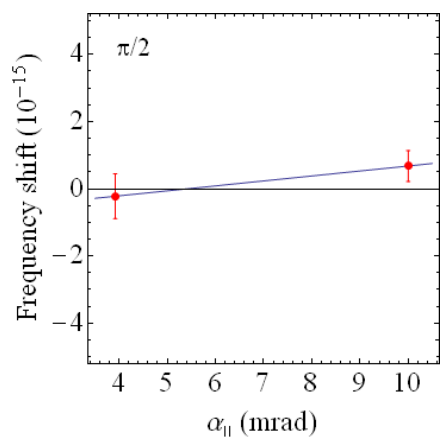

Figure 5. Dependence of the asymmetric DCP frequency on the relative angle of the fountain tilted away from vertical in the plane parallel to the feeds of the Ramsey cavity. The feeds of the cavity are balanced and correspond to a $\pi/2$ Ramsey resonance of the microwave field.

The resulting angular sensitivity of the fractional frequency of the Rb fountain frequency standard in the plane parallel to the feeds of the Ramsey cavity is equal to

$$\frac{1}{\nu}\frac{d\nu}{d\alpha_\|} = 1.4 \pm 1.3 \ (10^{-16}/\text{mrad}).$$

The corresponding angular sensitivity in the plane perpendicular to the feeds is equal to

$$\frac{1}{\nu}\frac{d\nu}{d\alpha_\perp} = 0.0 \pm 2.0 \ (10^{-16}/\text{mrad}).$$

The fit line in Fig. 6 is taken to have no slope, which corresponds to an absence of phase gradient in that direction.

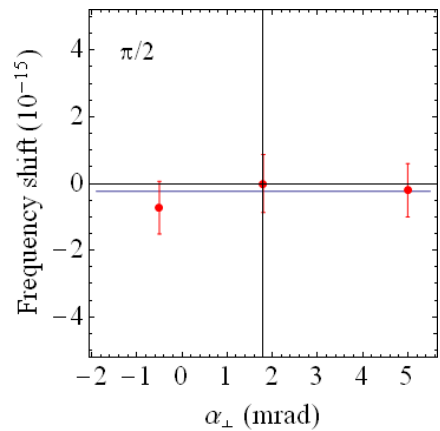

Figure 6. Dependence of the asymmetric DCP frequency on the relative angle of the fountain tilted away from vertical in the plane perpendicular to the feeds of the Ramsey cavity. The feeds of the cavity are balanced and correspond to a $\pi/2$ Ramsey resonance of the microwave field.

For an uncertainty in the vertical angle of the fountain of 0.2 mrad, the corresponding uncertainty in the asymmetric part (mainly the $m = 1$ term) of the DCP fractional frequency shift is $0.55\times 10^{-16}$ for the tilt of the fountain in the parallel plane and $0.4\times 10^{-16}$ for its tilt in the perpendicular plane.

### D. Estimate of the symmetric DCP-shift contributions

The symmetric DCP frequency shift consists mainly of the $m = 0$ and $m = 2$ terms, which are too small ($< 10^{-16}$) to be measured directly. The estimates of these shifts were based on mathematical modeling [9-12]. In our case we rely on the calculations [12] for the LNE-SYRTE Rb fountain, the Ramsey cavity of which has the same aspect ratio and size as our cavity. The only difference is that our cavity has a larger diameter of the cut-off tubes (16 mm instead of 12 mm) and a much higher quality factor (28500 instead of 6000).

The $m = 0$ contribution of the DCP frequency shift of the LNE-SYRTE Rb fountain [12] is estimated to be as small as $-0.2(5)\times 10^{-17}$. Therefore, it is absolutely safe in our case to take the uncertainty of the $m = 0$ DCP frequency shift to be equal to $0.3\times 10^{-16}$.

The estimate of the $m = 2$ DCP shift for the LNE-SYRTE Rb fountain, which includes the spatial inhomogeneity of the detection and a possible 2 mm transverse offset of the launched atoms, gives a total shift of $0.5(3) \times 10^{-16}$. Since the size of our initial atomic cloud (with radius of 3 mm at the 1/e density level) and the transverse inhomogeneity of the detection beam are very close to the LNE-SYRTE fountain, we take the uncertainty of the $m = 2$ frequency shift to be equal to $1.0 \times 10^{-16}$. In future we plan to turn the detection beams to be at 45° with respect to the feeds of the Ramsey cavity, as is done in the LNE-SYRTE Cs fountain FO1, to essentially suppress the $m = 2$ component of the DCP shift.

The total uncertainty budget for the DCP frequency shift of the NPL Rb fountain frequency standard is presented in Tab. I.

TABLE I. DCP-SHIFT UNCERTAINTY BUDGET

| Type of DCP shift | Uncertainty ($10^{-16}$) |
|---|---|
| $m = 1$ (parallel) | 0.55 |
| $m = 1$ (perpendicular) | 0.4 |
| $m = 0$ | 0.3 |
| $m = 2$ | 1.0 |
| Total | **1.25** |

III. MEASUREMENT OF ALL POWER DEPENDENT SHIFTS

Finally we have performed more detailed measurements of the dependence of the fountain frequency on the amplitude of the microwave field inside the Ramsey cavity. This time the measurements were done for the fountain precisely aligned vertically and after proper balancing of the feeds of the cavity, which minimizes the asymmetric DCP frequency shift.

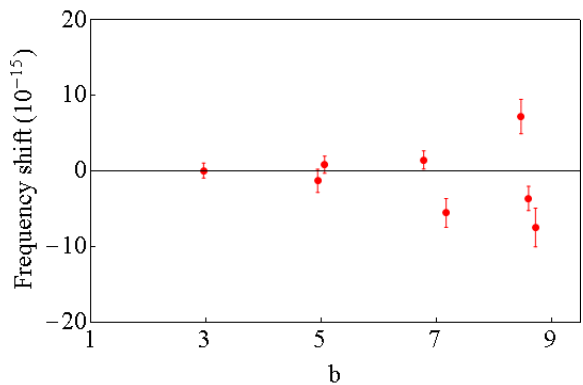

Figure 7. Dependence of the frequency shift of the fountain on the amplitude of the microwave field inside the Ramsey cavity, which is taken for a vertically aligned fountain and balanced feeds of the cavity.

This measurement was performed relative to the frequency shift at the optimal power (b=1). The results of these measurements are presented in Fig. 7.

This dependence includes all power dependent frequency shifts of the Rb fountain, the main contributions of which are the symmetric DCP shift and possible microwave leakage shift.

One can see that the frequency shifts turn to zero around all the Ramsey resonances of higher order (b=3, 5, 7, 9). Therefore, the general character of the observed dependence agrees well with the typical power dependence of the symmetric DCP frequency shift and indirectly confirms the smallness of the microwave leakage shift.

IV. SUMMARY

The DCP frequency shift of the NPL Rb fountain has been characterized and its total uncertainty is equal to $1.25 \times 10^{-16}$. The corresponding total type B uncertainty of the Rb fountain (Tab. 1) is reduced down to $2.4 \times 10^{-16}$.

TABLE II. TYPE B UNCERTAINTY BUDGET

| Effect | Frequency shift ($10^{-16}$) | Uncertainty ($10^{-16}$) |
|---|---|---|
| Second order Zeeman | 857.6 | 1.0 |
| Blackbody radiation | -133.5 | 0.6 |
| Distributed cavity phase | 0 | 1.25 |
| Collisions with residual gas | <0.7 | 0.7 |
| Microwave lensing | -0.9 | 0.5 |
| Microwave leakage | 0 | 1.0 |
| Gravity | 12.6 | 0.3 |
| Other | 0 | 1.0 |
| **Total** | | **2.4** |

Based on this improved accuracy of the NPL Rb fountain frequency standard we will perform a new absolute measurement of the Rb ground state hyperfine splitting.

ACKNOWLEDGMENT

I would like to thank Krzysztof Szymaniec and Rachel Godun for useful discussions of the results presented in this paper.